\documentclass[preprint,eqseqnum,12pt]{elsarticle}

\usepackage{epsfig}
\usepackage{amssymb}
\usepackage{amsmath}
\usepackage{lineno}
\biboptions{sort&compress}

\usepackage{bm}
\usepackage{amsmath}
\usepackage[utf8]{inputenc}

\numberwithin{equation}{section}
\journal{}

\begin{document}

\begin{frontmatter}




\title{The Mass Gap Approach to QCD. I. The true gauge and dynamical structures of its ground state }


\author{V. Gogokhia}
\ead{gogohia.vahtang@wigner.hu}
\author{G.G. Barnaf\"oldi}
\ead{barnafoldi.gergely@wigner.hu}

\address{Wigner Research Centre for Physics,\\

29-33 Konkoly-Thege Mikl\'os Str., H- 121, Budapest, Hungary}

\begin{abstract}
Assuming that a non-trivial quantum Yang-Mills theory exists,  we have proved that it should have a mass gap $\Delta^2 > 0$, indeed. The proof is based on the derivation of the novel constraint 
on any solution to QCD. It has been exactly and uniquely derived in the framework of the Slavnov-Taylor identities for the gauge particles Green's functions (propagators),
involving the equation of motion for the full gluon propagator as well. The novel constraint has the two different solutions, coinciding  only at high energies. 
The dynamical source of this difference has to be identified with the constant tadpole term, contributing to the full gluon self-energy. Just its renormalized version is conventionally called a mass gap. 
We prove that it cannot be disregarded from the theory and its ground state by any means.  The perturbative renormalizability of QCD will not be affected by a new solution for the gluon equation of motion.
We also  provide the formulation of the self-consistency condition for the gauge choice in QCD. Finally, we discuss the interrelation of our advance results with the Jaffe-Witten's theorem.
\end{abstract}



\end{frontmatter}




\section{Introduction}


The quark model (QM) treats the strongly-interacting particles (baryons and mesons)
as bound-states of quarks, emitting and absorbing gluons. The theory which should describe the properties of the observed hadrons in terms of the non-observable quarks and gluons from first principles is Quantum Chromodynamics (QCD)~\cite{1,2,3,4,5,6,7,8,9}. It is widely accepted as the quantum gauge field theory of strong interactions. Being such a theory, it has to undergo the two phase 
transitions: the first one at the fundamental (microscopic) quark-gluon level is -- the confinement phase transition -- in order to explain why all the physical states are colour-singlets.
The second one at the hadronic (macroscopic) level is -- the Partially Conserved Axial Currents (PCAC) phase transition -- in order to explain the soft pion physics~\cite{2,10}. 

However, this purpose remains a formidable task yet because of the multiple dynamical and topological complexities of low-energy particle physics, originated from QCD 
and its ground state (or, equivalently, vacuum). This happens because QCD as a gauge theory of the gluon-gluon and quark-gluon strong interactions still suffers from a few conceptual problems.
The $SU(3)$  colour gauge invariance of the QCD Lagrangian forbids any mass scale parameter apart from the currant quark mass term. So immediately arises the   
question how does a mass come out of a massless theory (the mass gap)~\cite{2} even at the fundamental quark-gluon level?
How to explain the colour confinement problem, i.e., why the coloured objects-- gluons and quarks-- cannot appear in the physical spectrum. It is an experimental fact and can be considered 
as the boundary condition at large-distances/low-energies regimes. 
The linear rasing potential between heavy quarks requires the mass scale parameter -- the string tension. A scale breaking in the asymptotic freedom (AF)~\cite{11,12} phenomenon
needs a mass squared scale parameter. It is also an experimental fact and can be considered as the boundary condition at  short-distances/high-energies regimes.
So that, the goal arises how to solve the above-mentioned  important problems, and thus to allow for QCD to explain confinement of the colour and the AF phenomenon at the same time.  
The solutions may come from the investigation of the true gauge and dynamical structures of the QCD vacuum in more detail. 

The properties and symmetries of the QCD Lagrangian, and thus including its Yang\,--\,Mills (YM) part, are well-known~\cite{1,2,3,4,5,6,7,8,9,13}.
The propagation of gluons is one of the main dynamical effects in the QCD ground state.
The importance of the corresponding equation of motion is due to the fact that its solutions are supposed to reflect the dynamical and gauge structures of the QCD ground state. 
The gluon Schwinger\,--\,Dyson (SD) equation~\cite{2,13,14,15,16,17,18,19,20,21,22,23} is a highly non-linear (NL) one because of the self-interaction of massless gluon modes, so the number of its independent solutions is not fixed {\sl a priori}. However, here we are going to investigate the structure of the QCD ground state with the help of the gluon equation of motion before
explicitly solving it. We will show that the general properties of the full gluon self-energy point out on some new dynamical and gauge aspects of the QCD ground state. 
Due to these inputs, the novel insights into the mass dynamical generation in the QCD vacuum are also present within proposed below formalism. It is based on the tensor algebra
derivation rigorous rules only, which is widely using in the theoretical and mathematical physics. 

Our primary aim is to solve the gluon confinement problem in the series, consisting of the three papers. The general title of this series is ''The mass gap approach to QCD'' since we already know the solution
of the just above-mentioned problem on the basis of the mass gap conception extended to the the confinement phase transition introduced previously in~\cite{13} (and see references therein)
and discussed in these papers. 
Here we have investigated the true gauge and dynamical structures  of the QCD ground state in more detail. The exact constraint on any solution to QCD has been derived.  
The proper gauge-invariant subtraction scheme has been formulated in order to remove from the theory the 
quadratic ultraviolet (UV) divergences, but without affecting its perturbation theory (PT) renormalizability. The new expression for the full gauge particle Green's function
(propagator) will be explicitly present. In the second paper the non-perturbative (NP) renormalization program for the massive gluon field configurations will be performed. 
The massive full gluon propagator always  will remain the off-mass-shell object at any finite gauge. The massive gluons cannot appear in the physical spectrum (confinement of the massive gluons). 
In the third paper the NP renormalization program for the singular gluon field configurations will be performed. The confining solution will depend on the 
transverse ''physical'' degrees of freedom in the full gluon propagator. The gluons will remain massless even in the explicit presence of a mass gap, but having the 
scale breaking AF behaviour at high energies. Only such solution will make it possible to calculate any observable in QCD from first principle, which is its ultimate goal, emphasized just above. 

Concluding, let us note that the comprehensive discussion of the present status of the confinement problem can be found in~\cite{24}.


\section{The gluon SD equation}


 In general, the system of the above-mentioned  SD dynamical equations of motion and the corresponding Slavnov-Taylor (ST) identities~\cite{25,26} for the  gluon and quark Green's functions (propagators)
complements each other in QCD. Together they contain much more informations on the properties of the theory and its ground state than its Lagrangian can provide at all ~\cite{2}. 
If there is no place for the mass scale parameter different from the current quark mass in the 
Lagrangian of QCD, then the only place where it may explicitly appear is the QCD ground state, which gauge and dynamical structures are described just by this system of the corresponding SD equations
and ST identities. So that, the gluon SD equation is the place where a mass squared scale parameter can be generated.      
This equation has a rather complicated tensor structure because of its NL character. In this connection, let us remind that all the known interactions in nature (electroweak, gravitational and strong) 
are described by the gauge theories. The common mathematical language for them is the tensor algebra, which is only one used throughout this paper.That is why our work will be easily 
understood and checked its each every step by a broad spectrum of researches from the different areas of the theoretical and mathematical physics, and not only by the particles and fields big community. 

The gluon SD equation analytically looks like

\begin{equation}
D_{\mu\nu}(q) = D^0_{\mu\nu}(q) + D^0_{\mu\rho}(q) i \Pi_{\rho\sigma}(q; D) D_{\sigma\nu}(q),
\end{equation}
where $D_{\mu\nu}(q)$ and $D^0_{\mu\nu}(q)$ denote the full gluon propagator and its free counterpart, respectively. $\Pi_{\rho\sigma}(q; D)$ is the full gluon self-energy which depends on the full gluon propagator due to the non-abeian character of QCD.
Here and everywhere below we omit the colour group indices, for simplicity, because of their final factorization, for example $D^{ab}_{\mu\nu}(q) = D_{\mu\nu}(q)\delta^{ab}$. Eq.~(2.1) in terms of the corresponding skeleton loop diagrams is shown in Fig. 1.

\begin{figure}[h!]
\begin{center}
\includegraphics[width=10.0truecm]{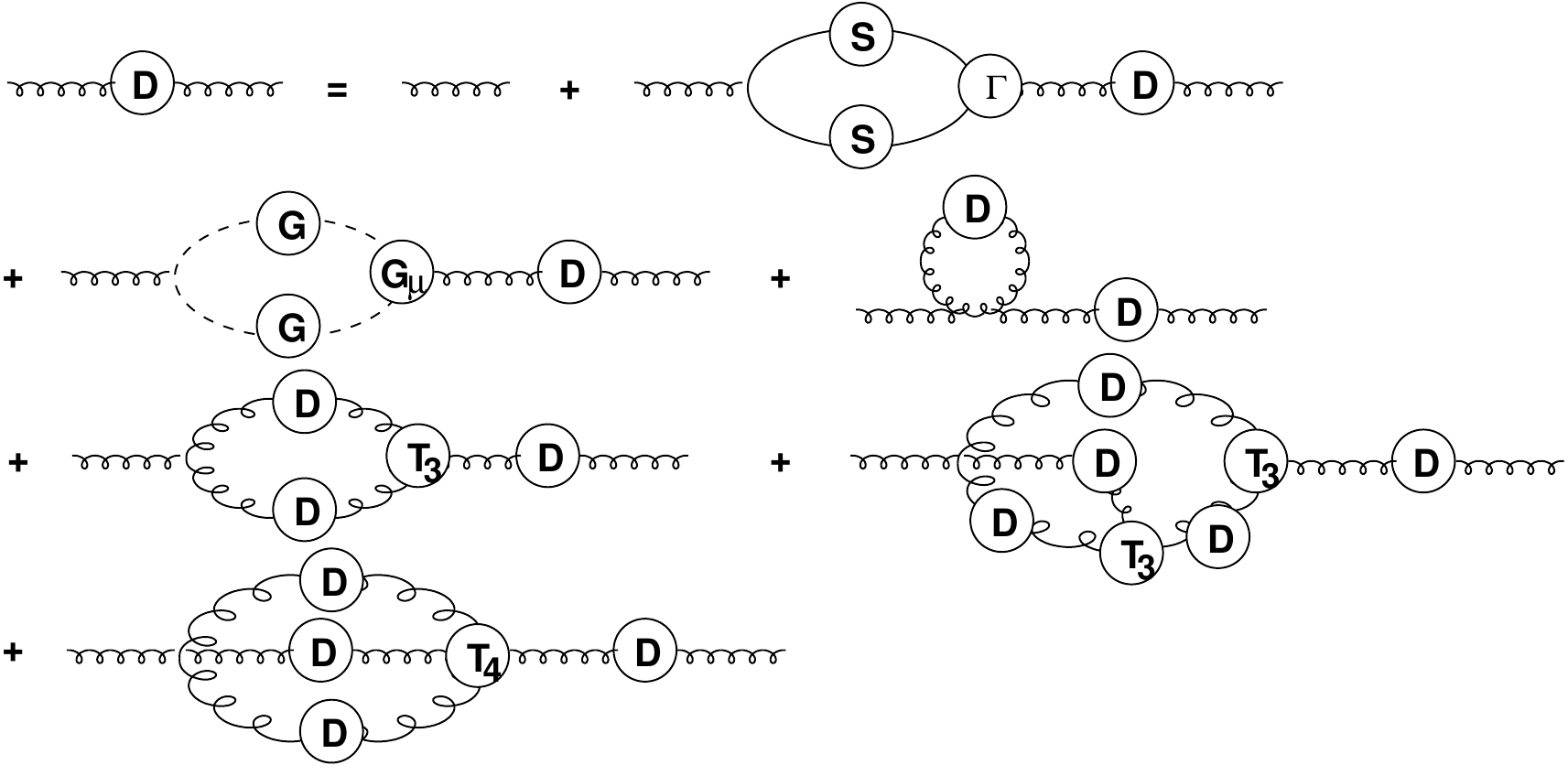}
\caption{The SDE for the full gluon propagator, as present in~\cite{13}.}
\label{fig:1}
\end{center}
\end{figure}

Here stringy lines are for the free gluon propagator, while $D$ denotes its full counterpart. $S$ with solid lines denotes the full quark propagator, and $\Gamma$ denotes the full quark-gluon vertex. $G$ with dashed lines denotes the full ghost propagator, and $G_{\mu}$ is
the full ghost-gluon vertex. Finally, $T_3$ and $T_4$ denote the full 3- and 4-gluon vertices, respectively.

The full gluon self-energy is convenient to present as the sum of the three independent terms, namely

\begin{equation}
\Pi_{\rho\sigma}(q; D) = \Pi^q_{\rho\sigma}(q) + \Pi^g_{\rho\sigma}(q; D) + \Pi_{\rho\sigma}^t(D),
\end{equation}
where $\Pi^q_{\rho\sigma}(q)$ describes the skeleton loop contribution of the quark degrees of freedom
as an analogue to the vacuum polarization tensor in Quantum Electrodynamics (QED)~\cite{7,27}.
Here and below the superscript or subscript '$q$' means quark (not to be mixed up with the gluon momentum variable $q$). The
gluon part of the full gluon self-energy by itself is the sum of a few independent terms as follows:

\begin{equation}
\hspace{-10mm} \Pi^g_{\rho\sigma}(q; D) = \Pi^{gh}_{\rho\sigma}(q) + \Pi^{(1)}_{\rho\sigma}(q; D^2) + \Pi^{(2)}_{\rho\sigma}(q; D^4) + \Pi^{(2')}_{\rho\sigma}(q; D^3),
\end{equation}
and $\Pi^{gh}_{\rho\sigma}(q)$ describes the skeleton loop contribution associated with the ghost degrees of freedom.
$\Pi^{(1)}_{\rho\sigma}(q; D^2)$ represents the skeleton loop contribution,
containing the 3-gluon vertices only. Finally, $\Pi^{(2)}_{\rho\sigma}(q; D^4)$ and $\Pi^{(2')}_{\rho\sigma}(q; D^3)$ describe the skeleton two-loop contributions, which combine the 3- and 4-gluon vertices. All these quantities are given by the corresponding skeleton loop diagrams in Fig. 1, and they are independent from each other.
The last four terms explicitly contain the full gluon propagators in the corresponding powers symbolically shown above.
The analytical expressions for the corresponding skeleton loop integrals~\cite{28}, in
which the symmetry and combinatorial coefficients and signs have been included, are not important here. We are not going to calculate any of them
explicitly, and thus to introduce into them any truncations/approximations/assumptions or choose some special gauge.
Any skeleton loop integral in Fig. 1 is a sum of the infinite number of terms. Moreover, the full
vertices entering these skeleton loop integrals are themselves determined by an infinite series of the corresponding multi-loop skeleton
terms~\cite{2,13,14,15,16,17,18,19,20,21,22,23} (see references therein). In these NL series the dependence on the coupling constant may not be simple, i.e., in fact,  
it is not known. Such kind of series are the so-called ''cluster'' expansions~\cite{10}.

The constant tadpole term $\Pi_{\rho\sigma}^t(D)$ is defined as, 

\begin{equation}
\Pi_{\rho\sigma}^t(D) \sim \int d^4 l D_{\alpha\beta}(l) T^0_{\rho\sigma\alpha\beta} = g_{\rho\sigma} \Delta^2_t(D),
\end{equation}
where $ g_{\rho\sigma} = T_{\rho\sigma} (q) + L_{\rho\sigma}(q) = T_{\rho\sigma} (q) +q_{\rho} q_{\sigma} / q^2$.
In what follows the constant $\Delta^2_t(D)$ will be called as the tadpole term itself, since it deviates from the definition (2.4) by the tensor $g_{\rho\sigma}$ only.   
So that, the tadpole term equally contributes to the transverse and longitudinal components of the full gluon self-energy (2.2).

All the terms which contribute to the full gluon self-energy eq.~(2.2) are tensors, having the dimensions of a mass squared. All these skeleton loop integrals are therefore
quadratically divergent (QD), i.e., UV divergent in the PT regime, and so they are assumed to be regularized from above.
Contrary to QED, QCD being a non-abelian gauge theory can suffer from the severe infrared (IR) singularities in the $q^2 \rightarrow 0$ limit (i.e., more singular than the PT one
$\sim 1/q^2$). Evidently, they will be generated due to the self-interaction of massless gluon modes. Thus, all the skeleton loop integrals, contributing to the full gluon self-energy, are assumed 
to be regularized from below as well. This means that all the expressions are regularized (i.e., we render infinite skeleton loop integrals to be finite). How precisely the regulating parameters 
have to be introduced is not important here, since we are not going to calculate any of these skeleton loop integrals explicitly, as emphasized above. 
They will be assumed but not explicitly shown in all the regularized expressions, for simplicity.

The constant tadpole term $\Delta^2_t(D)$, defined in eq.~(2.4), is nothing else but the QD, i.e, the UV divergent skeleton loop integral, which is already regularized from below and above, as well as all the other such kind of constants which may appear in the theory by any possible ways (the subtraction schemes, the iteration solutions, etc.). Being a mass squared regularized quantity, it is explicitly present  in the QCD ground state. Then a natural question arises why is it present in the vacuum of QCD at all, if it makes the theory to look non-renormalizable from the very beginning? 
Let us remind that the theory possessing the QD quantities is not the PT renormalizable one since the logarithmic-type divergences only can be accounted for the PT. 
The standard solution was to remove this and all other QD constants in any possible gauge-invariant way from the theory in order to make it renormalizable in the PT sense.
However, here we are going to ask the question is it possible to retain the tadpole term in the QCD vacuum, but without 
affecting the PT renormalizability of the theory? This is important to ask because, unlike to other QD constants, the tadpole term is only one which dynamically generates a mass squared scale parameter, dominated by the point-like four-gluon vertex only, and thus does not depending on the external gluon momentum at all~\cite{6}.
Also, its role in the dynamical and gauge structures of the QCD ground state has been clarified in more detail (see sections below). Being a mass squared scale parameter it violets the gauge symmetry of the
QCD Lagrangian in its true ground state from the very beginning. However, this is not so important, since in our second and third papers, we will show that the solutions  of a new full gluon propagator
will not depend on the  gauge choice, as mentioned above at the end of the previous section.  

Concluding, the whole gluon momentum range is $q^2 \in [0, \infty)$. In this paper we worked in Euclidean metric $q^2=q^2_0 + \bf q^2$ since
it implies $q_i \rightarrow 0$ when $q^2 \rightarrow 0$ and {\it vice versa}. This makes it possible to avoid the unphysical IR singularities on the light cone.

\section{Exact constraint on any solution to QCD}

It is well known that all the ST identities which appear in QCD are important for the renormalizability of the theory,
and thus they {\it {''are exact constraints on any solution to QCD''}}~\cite{2}.  Among them the ST identities~\cite{2,25,26}  for the gauge particle propagators (Green's functions) play a dominant role,
since they determine the gauge structure of the QCD ground state.  

The ST identity for the full gluon propagator looks like
\begin{equation}
q_{\mu}q_{\nu} D_{\mu\nu}(q) = i \xi,
\end{equation}
where $\xi$ is the gauge-fixing parameter. It implies that the general tensor decomposition of the full gluon propagator in the covariant gauge is as follows:
\begin{equation}
D_{\mu\nu}(q) = i \left[ T_{\mu\nu}(q) d(q^2) + \xi L_{\mu\nu}(q) \right] {1 \over q^2},
\end{equation}
where the invariant function $d(q^2)$ is the corresponding Lorentz structure of the full gluon propagator (the gluon invariant function). Throughout this paper we use the standard 
definition of $T_{\mu\nu}(q) = \delta_{\mu\nu} - q_{\mu}q_{\nu} / q^2= \delta_{\mu\nu} - L_{\mu\nu}(q)$ in Euclidean metric.
Any invariant functions associated with the projection operators $T_{\mu\nu}(q)$ and $\delta_{\mu\nu}$ are the same, and thus $D_{\mu\nu}(q)$ is defined up to its longitudinal 
part  $L_{\mu\nu}(q)$. This should be also valid for its equation of motion.

By formally setting $d(q^2)=1$ in eq.~(3.2), then one obtains the free gluon propagator
\begin{equation}
D^0_{\mu\nu}(q) = i \left[ T_{\mu\nu}(q) + \xi_0 L_{\mu\nu}(q) \right] {1 \over q^2},
\end{equation}
where $\xi_0$ is the corresponding gauge-fixing parameter. The general ST identity (3.1) will look like
\begin{equation}
q_{\mu}q_{\nu} D^0_{\mu\nu}(q) = i \xi_0.
\end{equation}
It is important to note in advance that from the very beginning  the gauge-fixing parameter for the full gluon propagator $\xi$ is not to be identified with its free counterpart $\xi_0$.
This possibility appears because the gauge-fixing action term is not present in the gluon equation of motion. Since  the gauge freedom in QCD is determined by $\xi_0$, there should 
exist a relation between them. 

Let us now perform some exact algebraic derivations which are necessary for the further purposes.
Contracting the full gluon SD eq.~(2.1) with $q_{\mu}$ and $q_{\nu}$, on account  of the relations (3.1)-(3.4), one gets

\begin{equation}
q_{\rho}q_{\sigma} \Pi_{\rho\sigma}(q; D)  = {( \xi_0 - \xi) \over \xi \xi_0 } (q^2)^2, 
\end{equation}
where the full gluon self-energy, due to the eqs.~(2.2) and (2.4), can be present as follows:

\begin{equation}
\Pi_{\rho\sigma}(q; D) =  \Pi^q_{\rho\sigma}(q)  + \Pi^g_{\rho\sigma}(q; D) + \delta_{\rho\sigma} \Delta^2_t(D),
\end{equation}
and contracting it with with $q_{\rho}$ and $q_{\sigma}$, and because of the previous relation (3.5), one  obtains 

\begin{equation}
q_{\rho} q_{\sigma} [ \Pi^q_{\rho\sigma}(q) + \Pi^g_{\rho\sigma}(q; D)  ]  = {( \xi_0 - \xi) \over \xi \xi_0 } (q^2)^2 - q^2 \Delta^2_t(D).
\end{equation}
It explicitly depends on the tadpole term $\Delta^2_t(D)$. 
If it is formally put zero, i.e., by hand, then the both transverse relations (3.5) and (3.7) coincide with each other. This underlines its important role in the QCD ground state, which 
one precisely is to be clarified in our paper.  
The transverse relation (3.5) and the transverse relation obtained from eq.~(3.6) are independent from each other, since the first one has been derived by 
using the gluon SD eq.~(2.1), while the second one by using the initial definition (2.2).  
From the transverse relations one concludes that by themselves they cannot remove the QD constants or, equivalently, the UV divergences, which may appear in the theory 
(in order to make it in the PT sense  renormalizable). Also, it will be not possible  to fix the relation between $\xi$ and  $\xi_0$. 

Let us now remind that in QCD the quark skeleton loop contribution into the full gluon self-energy can be always made transverse because of the electric
charge conservation which flows around the closed loop (see Figure 1 and its description in the text above) in complete analogy with the vacuum polarization tensor
in QED~\cite{7,27}. Therefore, $q_{\rho} q_{\sigma} \Pi^q_{\rho\sigma} = 0$ 
and this relation holds independently  from the gluon contribution, which has been defined in eq.~(2.3). At the same time,
it is well-known that in QCD just the Faddeev\,--\,Popov (FP) ghost contribution~\cite{29} $\Pi^{gh}_{\rho\sigma}(q)$ makes the transverse relation for this term satisfied
(independently from the tadpole term (2.4), which itself has the transverse projection operator structure).
It is instructive to emphasize that each term in the sum (2.3) cannot be made transverse, only the whole sum. How explicitly this works can be found in any text book on QCD. 
For the most transparent calculations up to one loop contribution to the transverse gluon self-energy (2.3) see for example~\cite{3,4,5,6}.  Of course, such a transversality will be achieved  
in the higher loop iteration terms as well, i.,e, it is a general fact, as it is requested by the satisfied transverse relation for the (2.3) term. It is a sum of the skeleton loop contributions. 
This means that $q_{\rho} q_{\sigma} \Pi^g_{\rho\sigma} (q; D) = 0$ at any number of the iteration loops, and this relation holds independently from the quark contribution. 

So that, in general, in QCD one has

\begin{equation}
q_{\rho} q_{\sigma} \Pi^q_{\rho\sigma} (q) = q_{\rho} q_{\sigma} \Pi^g_{\rho\sigma}(q; D)  =  0,   
\end{equation}
and then from the relation (3.7), one immediately arrives at the exact condition, namely

\begin{equation}
{( \xi_0 - \xi) \over \xi \xi_0 } q^2 =\Delta^2_t(D), 
\end{equation}
which can be treated as the exact constraint on any solution to QCD, since coming out from the corresponding ST identities, and
deriving before going to the formulation of any proper subtraction scheme.This underlines its intrinsic, independent and important status in the theory. 

\hspace{1mm}

{\bf NL solution:} The general solution of the exact constraint (3.9) determines the function $\xi = f(q^2; \xi_0), \Delta^2_t(D))$ as follows:

\begin{equation}
\xi = f(q^2; \xi_0, \Delta^2_t(D))  = {  \xi_0  q^2 \over  q^2  + \xi_0 \Delta^2_t(D)}.                   
\end{equation}
This relation has been exactly (i.e., without making any kind of the simplifications) and uniquely defined in a such 
new manner for the first time.This expression demonstrates the NL dependence $\xi = f(q^2; \xi_0), \Delta^2_t(D))$ on $\xi_0$ and $\Delta^2_t(D)$.
The linear relation $\xi \rightarrow \xi_0$ will be recovered in the PT $q^2 \rightarrow \infty$ limit, when the ratio $(\Delta^2_t(D) / q^2)$ is to be suppressed
in this regime at finite $\xi_0$.  We have established the boundary condition for the behaviour of this relation in the PT limit for the regularized full gluon propagator.

\hspace{1mm}

{\bf Linear solution:} The exact constraint (3.9) has also a particular solution, namely 

\begin{equation}
\Delta^2_t(D) = 0 \ \rightarrow \ \xi=\xi_0, \ and \ {\it vice \ versa}, 
\end{equation}
but these equalities should be put by hand, being thus the prescriptions.


\section{Proper subtraction scheme}


QCD being much more complicated quantum field gauge theory than QED~\cite{7,27} has the two independent satisfied transverse relations (3.8), and 
so requires much more careful investigation in QCD. They have to be included in a self-consistent way into the proper subtraction scheme. Finally this will lead to the removal of the 
corresponding QD constants from the theory. How the corresponding gauge-fixing parameters have been determined  via the relations (3.10) and (3.11) has been just described above. 

The first step in the renormalization program of any gauge theory is the removal of the quadratic  UV divergences in order to make
the corresponding theory renormalizable in the PT sense. It can be achieved by introducing the proper subtraction scheme in order to separate them from the PT logarithmic divergences. The preliminary 
step in the regularization program has been already done by introducing the corresponding regulating parameters, mentioned in section 2.
Within our approach nothing will depend on how exactly the regulating parameters have been introduced. Even the hard cut-off procedures can be used, since we are mainly interested 
in the regulating of leading order UV divergences~\cite{2}. They have to disappear from the theory after the PT and the NP renormalization programs will be performed. 

The subtraction of  the quark contribution to the full gluon self-energy is to be defined  as follows: $\Pi^q_{\rho\sigma}(q)= \Pi^q_{\rho\sigma}(q) - \Pi^q_{\rho\sigma}(0) + \Pi^q_{\rho\sigma}(0)$, i.e., 
we adding zero to the corresponding identity, i.e., they have not been introduced by hand. The initial quark contribution to the gluon self-energy is not changed and thus is being gauge-invariant.
The initial quark contribution then becomes $\Pi^q_{\rho\sigma}(q)= \Pi^{s(q)}_{\rho\sigma}(q) + \Pi^q_{\rho\sigma}(0)$ as a sum of the two terms, one of which shows up all the corresponding QD but regularized constants.The first term is zero when the external gluon momentum $q=0$, by definition, and it may be only logarithmically divergent at large $q$. 
In other words, we separate the NP QD constants from the PT logarithmically divergent terms. The same is true for the initial gluon contribution as well.
Such an exact separation will be very useful for the renormalization programs of any kind to be performed for the single full gluon propagator. For the interacting full gluon propagators these independent terms will interact with each other, but such a separation will simplify the corresponding renormalization programs, anyway.

Let us now explicitly proceed to  the subtractions for the quark and gluon contributions to the full gluon self-energy as follows:

\begin{equation}
\Pi^{s(q)}_{\rho\sigma}(q) = \Pi^q_{\rho\sigma}(q) - \Pi^q_{\rho\sigma}(0) =   \Pi^q_{\rho\sigma}(q) - \delta_{\rho\sigma}\Delta^2_q, 
\end{equation}
and

\begin{equation}
\Pi^{s(g)}_{\rho\sigma}(q; D) = \Pi^g_{\rho\sigma}(q; D) - \Pi^g_{\rho\sigma}(0; D) = \Pi^g_{\rho\sigma}(q; D) - \delta_{\rho\sigma}\Delta^2_g(D), 
\end{equation}
where and everywhere below the superscript ''s'' means that the subtractions are already made, and thus $\Pi^{s(q)}_{\rho\sigma}(0) = \Pi^{s(g)}_{\rho\sigma}(0; D)  = 0$, by definitions. 
In these relations $\Delta^2_q$  is the skeleton loop integral for the quark degrees of freedom at $q=0$, i.e., $\Pi^q_{\rho\sigma}(0) = \delta_{\rho\sigma}\Delta^2_q$.
In the same way, $\Delta^2_g(D)$ is the sum of the corresponding skeleton loop integrals at $q=0$ contributing to eq.~(2.3), i.e., $\Pi^g_{\rho\sigma}(0; D) = \delta_{\rho\sigma} \Delta^2_g(D)$,
namely

\begin{equation}
\Delta^2_g(D) = \Delta^2_{gh} + \Delta^2_1(D^2) + \Delta^2_2(D^4) + \Delta^2_{2'}(D^3).
\end{equation}

It is worth reminding that all the QD constants, shown up in eqs.~(4.1)-(4.3), as well as the tadpole term $\Delta^2_t(D)$ itself, are independent from each other and are 
defined by the skeleton loop integrals, which have been already regularized from above and below.
The subtraction at zero is to be understood in a such way that we subtract at the external gluon momentum $q^2 = - \mu^2$~\cite{2} with $\mu^2 \rightarrow 0$ final limit. 
However, due to the self-interaction of the massless gluon modes, some of these constants (apart from the tadpole term) may depend on the internal zero gluon momentum 
(if it is connected to any closed loop,  which can appear in the iteration solution of the gluon SD equation~\cite{6}, and see appendix A as well). 
Then it has to be replaced by the subtraction at $q^2_i = - M^2_i$ with $M^2_i \rightarrow 0$ final  
limit, and subindex $i$ determines the number of such internal gluons.  For the any single full gluon propagator the subindex $i$  can be treated  as the number of
the necessary subtractions made in it in accordance with the rules of the theory of distributions (the generalized functions)~\cite{30}.

The independent tensor decompositions of the quark and gluon degrees of freedom, which appear in the subtraction relations (4.1)-(4.2), are 

\begin{eqnarray}
\Pi^{s(q)}_{\rho\sigma}(q) &=&  T_{\rho\sigma}(q) q^2 \Pi^{s(q)}_t(q^2) - q_{\rho} q_{\sigma} \Pi^{s(q)}_l(q^2),    \nonumber\\
\Pi^q_{\rho\sigma}(q) &=&   T_{\rho\sigma}(q) q^2 \Pi^q_t(q^2) - q_{\rho} q_{\sigma} \Pi^q_l(q^2),       
\end{eqnarray}
and

\begin{eqnarray}
\Pi^{s(g)}_{\rho\sigma}(q) &=&  T_{\rho\sigma}(q) q^2 \Pi^{s(g)}_t(q^2) - q_{\rho} q_{\sigma} \Pi^{s(g)}_l(q^2),    \nonumber\\
\Pi^g_{\rho\sigma}(q) &=&   T_{\rho\sigma}(q) q^2 \Pi^g_t(q^2) - q_{\rho} q_{\sigma} \Pi^g_l(q^2),       
\end{eqnarray}
respectively. In all the quantities above and below the dependence on $D$ is omitted, for simplicity, and will be restored when necessary.
Here and everywhere below all the invariant functions are dimensionless ones of their argument $q^2$: otherwise they remain arbitrary. However,
all the invariant functions $\Pi^{s(q)}_t(q^2)$, $\Pi^{s(q)}_l(q^2)$ as well as $\Pi^{s(g)}_t(q^2)$, $\Pi^{s(g)}_l(q^2)$
cannot have the power-type singularities (or, equivalently, the pole-type or the massless ones)  at small $q^2$, by definitions, i.e., their are regular functions of their arguments, as it follows from the initial subtractions (4.1)-(4.2).

Substituting all these decompositions (4.4)-(4.5) into the subtractions (4.1)-(4.2), and doing some tensor algebra derivations, one finally obtains

\begin{equation}
\Pi^{s(q)}_t(q^2) = \Pi^q_t(q^2)  - {\Delta^2_q \over q^2},  \quad \Pi^{s(q)}_l(q^2) = \Pi^q_l(q^2)  + {\Delta^2_q \over q^2},
\end{equation}
and

\begin{equation}
\Pi^{s(g)}_t(q^2) = \Pi^g_t(q^2)  - {\Delta^2_g(D) \over q^2},  \quad \Pi^{s(g)}_l(q^2) = \Pi^g_l(q^2)  + {\Delta^2_g(D) \over q^2}.
\end{equation}
Using these relations it is easy to show that  for the regularized (i.e., finite) quark and gluon contributions to the initial gluon self-energy their values at zero always are
$\Pi^q_{\rho\sigma}(0; D) = \delta_{\rho\sigma} \Delta^2_q$ and $\Pi^g_{\rho\sigma}(0; D) = \delta_{\rho\sigma} \Delta^2_g(D)$, indeed,
i.e, they are general ones, and have been already used in the subtraction relations (4.1)-(4.2) above, as it has to be.  Also, these relations will be true if expressed in terms 
of the above-mentioned subtraction point $q^2 = - \mu^2$.  

An interesting observation follows from the relations (4.6) and (4.7), namely 
$\Pi^q_t(q^2)  +  \Pi^q_l(q^2) = \Pi^{s(q)}_t(q^2)  + \Pi^{s(q)}_l(q^2)$ and 
$\Pi^g_t(q^2)  +  \Pi^g_l(q^2) = \Pi^{s(g)}_t(q^2)  + \Pi^{s(g)}_l(q^2)$.
So that the corresponding sums do not depend on the corresponding QD but regularized constants, which once more underlines the general character of our subtraction scheme.
The sum of the subtracted invariant functions is never zero, though the subtracted contributions to the full gluon self-energy are zero at $q^2=0$, by definitions, see relations (4.1)-(4.2).

As we already know, the quark and gluon contributions to the full gluon self-energy are transverse $q_{\rho} q_{\sigma} \Pi^q_{\rho\sigma}(q) = q_{\rho} q_{\sigma} \Pi^g_{\rho\sigma}(q; D) = 0$,
see the relations (3.8). Then from the second of the relations (4.6) and (4.7) it follows that  $\Pi^q_l(q^2) = \Pi^g_l(q^2) = 0$ as well. So that they will be reduced to  

\begin{equation}
\Pi^{s(q)}_l(q^2) = {\Delta^2_q \over q^2},  \quad  \Pi^{s(g)}_l(q^2; D)  = {\Delta^2_g(D) \over q^2},
\end{equation}
where we restored the dependence on $D$ in the second relation. However, these relations are impossible,
since the corresponding invariant functions cannot have the pole-type singularities, by definitions, as explained above.
One has to put these constants to zero on this general mathematical basis, i.e., $\Delta^2_q = \Delta^2_g(D) =0$, and thus
$\Pi^{s(q)}_l(q^2)  = \Pi^{s(g)}_l(q^2; D) =0$ as well.  

All our results for the QD constants can be now summarized as follows: 

\begin{equation}
\Delta^2_q = \Delta^2_g(D) =0, \quad \Delta^2_t(D) \neq 0,
\end{equation}
and hence $\Delta^2_{gh} = \Delta^2_1(D^2) = \Delta^2_2(D^4) = \Delta^2_{2'}(D^3) = 0$,
as well because of the relation (3.4). The system of the relations (4.9) is not only a general but it is a unique one as well.  They are not prescriptions, since  based on
a rigorous tensor algebra derivation rules, i.e., they are exact mathematical results. That the constant tadpole term $\Delta_t^2(D)$ remains intact is a general feature of any 
subtraction scheme. Constant minus the same constant is always zero, and it is a unique one as well in the sense that only the tadpole term $\Delta_t^2(D)$  will appear in the final regularized expressions
(see below). The essential dynamical source of its survival in the theory is that only the tadpole term can generate a mass squared scale parameter~\cite{6}, because of not depending 
on the external gluon momentum $q$. All the other quark and gluon terms, contributing to the full gluon self-energy, cannot do this, and thus none of their subtracted counterparts 
$\Delta^2_q$ and  $\Delta^2_g(D)$ can be treated as the mass gaps.

All these constants, defined by the corresponding skeleton loop integrals at $q=0$, are QD at the upper limit. Let us remind that each skeleton integral is a sum of the infinite number 
of the corresponding terms, as underlined in section 2. They have to be removed/disregarded from 
the theory on the general mathematical basis, i.e., put zero, as described above, apart from the tadpole term (2.4). In comparison with it, all the other QD quark and gluon constants will 
be called the tadpole-like/type terms. The general question arises now, namely how to understand these exact equalities to zero in the relations (4.9)? 
These equalities mean that any tadpole-like term which may appear in the theory by any possible way has to be discarded/disregarded in the theory, i.e, put zero,
independently from any other tadpole-like terms. In this way such sums will be always zero, indeed. 
Such constants may appear even not as the result of the subtractions, but, for example as the result of the NL iteration procedure for the full gluon propagator~\cite{6}.
In this case the tadpole-like terms may be even multiplied by some regularized functions (see appendix A). Whatever their origins would have been in the full gluon 
propagator, all of them belong to the infinite manifold of the relations (4.9), and thus should be always removed from the theory.

 Collecting our results obtained above for the quark and gluon contributions to the gluon self-energy, one obtains
 
\begin{equation}
\Pi^q_{\rho\sigma}(q) = T_{\rho\sigma}(q) q^2   \Pi^{s(q)}_t(q^2), \quad \Pi^g_{\rho\sigma}(q; D) = T_{\rho\sigma}(q) q^2   \Pi^{s(g)}_t(q^2; D),
\end{equation}   
since they become transverse and coincide with their subtracted counterparts. 

Substituting further the sum of these two terms into the full gluon self-energy (3.6), it becomes

\begin{equation}
\Pi_{\rho\sigma}(q; D) = T_{\rho\sigma}(q) q^2   \Pi^s (q^2; D) + \delta_{\rho\sigma} \Delta^2_t(D),
\end{equation} 
where $\Pi^s (q^2; D)  =\Pi^{s(q)}_t(q^2) + \Pi^{s(g)}_t(q^2; D)$,
reminding also that the invariant functions present in the previous relations are regular functions 
at small $q^2$, i.e., they have no pole-type singularities, and may be only logarithmically divergent at large $q^2$. This becomes possible only due to the satisfied transverse relations 
(3.8) for the quark and gluon degrees of freedom. Just this decreases the quadratic UV divergences of the corresponding skeleton loop integrals to a logarithmic ones, as it has 
been described in this section.


\section{New solution to the gluon SD equation (2.1)}


Substituting eq.~ (4.11), on account of the relation $\delta_{\rho\sigma} = T_{\rho\sigma}(q) +  L_{\rho\sigma} (q)$, further 
into the full gluon SD eq.~(2.1) and doing some tensor algebra, it finally becomes

\begin{eqnarray}
D_{\mu\nu}(q) = D^0_{\mu\nu}(q) &+& D^0_{\mu\rho}(q)i T_{\rho\sigma}(q) [ q^2 \Pi^s(q^2; D) + \Delta^2_t(D) ] D_{\sigma\nu}(q)  \nonumber\\
&+& D^0_{\mu\rho}(q)i  L_{\rho\sigma} (q) \Delta^2_t(D) D_{\sigma\nu}(q).
\end{eqnarray} 
 
The invariant function $\Pi^s(q^2; D)$, defined just above, is regular at zero and may have only the logarithmic divergences 
in the PT $q^2 \rightarrow \infty$ limit. The gluon SD eq.~(5.1), being derived from the initial gluon SD eq.~(2.1), has been re-written in the different form. 
It is free from all the other QD constants, but the regularized tadpole term remains, as expected.   

Combining the gluon SD eq.~(5.1) with the decompositions (3.2) and (3.3), one obtains

\begin{equation}
d(q^2) = {1 \over 1 + \Pi^s(q^2; D) + (\Delta^2_t(D) / q^2)}.
\end{equation}
This relation is the NL transcendental equation for the different invariant functions $d(q^2), \ \Pi^s(q^2; D)$ and the constant $\Delta^2_t(D)$,
i.e., $d=f(D(d))$. Nevertheless, from this expression is clearly seen that in the PT $q^2 \rightarrow \infty$ regime, the contribution
$(\Delta^2_t(D) / q^2)$ can be neglected, but the invariant function $\Pi^s(q^2; D)$ may still depend on this ratio under the PT logarithms.
In the NP region of finite and small gluon momenta this term is dominant, and the dependence of $d(q^2)$ on $(\Delta^2_t(D) / q^2)$ 
may be much more complicated due to the transcendental character of eq~(5.2). 

Contracting the full gluon SD eq.~(5.1) with $q_{\mu}$ and $q_{\nu}$, and substituting its result into the general ST identity (3.1), one arrives at
\begin{equation}
q_{\mu}q_{\nu} D_{\mu\nu}(q) = i \xi_0 \left( 1 - \xi  {\Delta^2_t(D) \over q^2} \right) =  i \xi,
\end{equation}
which solution is
\begin{equation}
\xi \equiv \xi(q^2; \xi_0, \Delta^2_t(D)) = { \xi_0 q^2 \over q^2 + \xi_0 \Delta^2_t(D)},
\end{equation}
i.e., in this case the gauge-fixing parameter becomes  $\xi \equiv \xi(q^2; \xi_0, \Delta^2_t(D))$ and thus it is not a constant equal to $\xi_0$.  
Let us point out that the expression (5.4) coincides with the general solution (3.10), which is as it should be.
So that, its asymptotic properties have been already discussed in the text after eq.~(3.10). The function (5.4) is the known function of its arguments, while
the relation (5.2) is the NL transcendental one. Behind the general inequality $\xi \neq \xi_0$ is the regularized constant $\Delta^2_t(D)$ as its dynamical source.  

Substituting eqs.~ (5.2) and (5.4) into the general decomposition (3.2) for the full gluon propagator, one finally obtains

\begin{equation}
D_{\mu\nu}(q) =   i T_{\mu\nu}(q) { 1 \over q^2 + q^2 \Pi^s(q^2; D) + \Delta^2_t(D) }  + i L_{\mu\nu}(q) { \lambda^{-1} \over q^2 + \lambda^{-1} \Delta^2_t (D)},
\end{equation}
where we have introduced the useful notation $\xi_0 = \lambda^{-1}$~\cite{4}. The corresponding ST identity now becomes

\begin{equation}
q_{\mu}q_{\nu} D_{\mu\nu}(q) = i \xi(q^2; \lambda^{-1}, \Delta^2_t(D))  = i { \lambda^{-1} q^2 \over q^2 + \lambda^{-1} \Delta^2_t(D)}.
\end{equation}
The ST identity (5.6) depends on the constant $\Delta^2_t(D)$, and when it is zero, one recovers the gauge-fixing parameter for the free gluon propagator.
In this ST identity  the gauge-fixing parameter $\lambda^{-1}$ for the free gluon propagator is convenient to vary continuously from zero to infinity.
The functional dependence of  $\xi$ (5.6) is fixed up to an arbitrary gauge-fixing parameter $\xi_0 = \lambda^{-1}$. Unless we fix it, and 
thus $\xi$ itself, we will call such situation as the generalized gauge dependence (GGD), see eqs.~(3.1)-(3.2) and (5.5)-(5.6).
Choosing $\xi_0 = \lambda^{-1}$ explicitly, we will call such situation as the explicit gauge dependence (EGD).
For example, $\xi_0 = \lambda^{-1}=0$ is called the unitary (Landau) gauge, $\xi_0 = \lambda^{-1}=1$ is called the t' Hooft--Feynman gauge, etc.~\cite{31,32,33,34}. 
The formal $\xi_0 = \lambda^{-1}= \infty$ limit is called as the canonical gauge in~\cite{33}. This distinction seems a mere convention, but, nevertheless, it is useful one in QCD because
of the presence of $\Delta^2_t(D)$ in its ground state. There is no other functional expression for $\xi$, apart from given by the relation (5.4) 
at finite  $\xi_0 = \lambda^{-1}$, in the full gluon propagator (5.5) and the ST identity (5.6) for the regularized gluon fields. 

The system of the regularized eqs.~(5.5)-(5.6), explicitly depending on the tadpole term, are present in the form suitable for the NP renormalization program to be performed.
The tadpole term enters the full gluon self-energy linearly, see the expression (3.6).
However, in the full gluon propagator (5.5) it appears in the NL way, because its contribution has been summed up with the help of the gluon SD eq.~(5.1).
The term $(\Delta^2_t(D)/ q^2)$ in the expressions (5.5)-(5.6)
is to be suppressed in the PT $q^2 \rightarrow \infty$  limit at the finite $\xi_0 = \lambda^{-1}$ (the above-mentioned canonical gauge will be investigated in detail in the forthcoming second paper).  
Then the full gluon propagator will behave like the free gluon propagator $\sim 1/ q^2$ in this limit.
This is the one of the necessary constraints that a theory is perturbative renormalizable. Other ones such as the corresponding behaviour of the spinor Green's function, a unitary of $S$-matrix 
and analyticity (causality)~\cite{2}, evidently are beyond the scope of the present work.  Here one can conclude that the true dynamical and gauge structures of the QCD/YM ground state 
are much more complicated than it follows from its Lagrangian's formulation. However, its PT renormalizability is not 
affected within the QCD full gluon propagator in the generalized gauge (5.6), i.e., it has the PT renormalizable behaviour at large $q^2$.

Concluding, it is worth noting that in terms of the transverse relations and the corresponding QD regularized constants as well as the corresponding relations between $\xi$ and $\xi_0$,
the new solution can be shown explicitly as the following system of the relations, namely

\begin{eqnarray}
\hspace{-10mm} q_{\rho} q_{\sigma} \Pi^q_{\rho\sigma} (q) &=& q_{\rho} q_{\sigma} \Pi^g_{\rho\sigma}(q; D)  =  0,  \nonumber\\ 
q_{\rho} q_{\sigma} \Pi_{\rho\sigma} (q; D) &=& q^2 \Delta^2_t(D),
\end{eqnarray}
and

\begin{equation}
\Delta^2_q = \Delta^2_g(D) =0, \ \Delta^2_t(D) \neq 0, \ \xi \neq \xi_0= \lambda^{-1}.
\end{equation}
It is necessary to underline that none of the prescriptions (i.e., putting some of them zero by hand)
have been introduced and none of the truncations, approximations and assumptions have been done, as well as no gauge fixing $\xi_0 = \lambda^{-1}$ by hand, in the obtaining of these relations. 
In other words, all of them are exact mathematical results in the solution  to the gluon SD equation.

Concluding, the system of the regularized eqs.~(5.5)-(5.6) represents a new solution to the initial gluon SD equation (2.1) in QCD.

\hspace{1mm}

\section{Renormalized ST identity and the mass gap}

\hspace{1mm}

Since the derivation of the novel constraint (3.9) has been mainly based on the ST identities formalism, it is useful to perform the NP renormalization program for the ST identity (5.6)
itself. It is instructive to present (5.6) in terms of $\xi$ and $\xi_0$ again, then we have (omitting the dependence of $\xi$ on  $\xi_0$ and $\Delta^2_t(D)$, for simplicity)

\begin{equation}
q_{\mu}q_{\nu} D_{\mu\nu}(q) = i \xi  = i {  \xi_0  q^2 \over  q^2  + \xi_0 \Delta^2_t(D)}.      
\end{equation}

Let us now introduce the NP renormalization constant for the tadpole term as follows: $\Delta^2_t(D) =  Z_{\Delta} \Delta^2$, where $\Delta^2$ is finite and positive, by
definition, while the NP renormalization constant $Z_{\Delta}$ depends on all the unphysical parameters, such as regulating parameters, mentioned above, etc. 
In principle, the string tension in the linear rising potential between heavy quarks has to be somehow related to the finite $\Delta^2$, since it is connected to the
transverse projection operator structure, as underlined above. It is well-known that the confining ansatz $ D_{\mu\nu}(q) \sim T_{\mu\nu}(q) \Delta^2 / (q^2)^2$ just leads to a such potential.
So that, the ST identity for the renormalized full gluon propagator becomes

\begin{equation}
q_{\mu}q_{\nu} D^R_{\mu\nu}(q) = i \xi^R  = i {  \tilde \xi_0  q^2 \over  q^2  + \tilde \xi_0 \Delta^2},      
\end{equation}

where the renormalized gauge-fixing parameters are defined as follows: $\tilde \xi_0 = Z_{\Delta} \xi_0, \ \xi^R = Z_{\Delta} \xi$ and 
$D^R_{\mu\nu}(q) =  Z_{\Delta} D_{\mu\nu}(q)$, i.e., the ST identity (6.2) is expressed in terms of the finite quantities only. The correct PT limit 
is maintained, of course. {\bf The renormalized version of the tadpole term $\Delta^2$ can be considered as a scale determining the NP structure of the QCD theory and its true vacuum.
Conventionally, we call it as a mass gap}. Such a finite mass gap  $\Delta^2$ will appear in the solution for the full gluon propagator as well (see the third paper of our series).

\hspace{1mm}

\subsection{The self-consistency condition for the gauge choice in QCD}

\hspace{1mm}

Let us now demonstrate the interesting feature of the generalized gauge (6.2). It provides the self-consistency condition for the gauge choice in QCD present as follows:

\begin{equation} 
{ \tilde \xi_0  q^2 \over  q^2  + \tilde \xi_0  \Delta^2} =  { a  q^2 \over  q^2  + a  \Delta^2},      
\end{equation}
i.e., the left-hand-side of this equation is present by the generalized gauge expression (6.2), while its right-hand-side presents the same expression
when the gauge is already chosen. In other words, we are checking whether the above-mentioned GGD formalism is compatible with its EGD  
one and $\it vice \ versa$. So that its aim is to derive a relation (not an identity) involving the gauge-fixing parameter.
If $a$ is any finite number, then from the self-consistency relation (6.3) it is easy to derive that $ \tilde \xi_0 = a$, indeed.

At the same time, if $a = \infty$ the so-called canonical gauge~\cite{33}, then the relation (6.3) becomes

\begin{equation}
{\tilde \xi_0 q^2 \over q^2 +  \tilde \xi_0 \Delta^2} =  {q^2 \over \Delta^2},
\end{equation}
which is only satisfied at $q^2=0$, i.e., there is no any condition for the gauge-fixing parameter. 
The formulated self-consistency condition (6.3) points out  on the inconsistency of the canonical gauge $\tilde \xi_0 = \infty$ in QCD. 
If everything is expressed in the terms of the finite quantities then the formal canonical gauge is forbidden to use in QCD, due to our approach to this theory. 

The self-consistency condition has been formulated not as some kind of the parametrization, but follows from the exact solution of the unchanged ST identities (3.1) and (5.3). 
As underlined above, they are important for the renormalizability of the theory. That is why the inconsistency of the canonical gauge is a physically meaningful.
Its physical inconsistency  will be discussed in detail in the second paper of our series.


\section{Previous solution to the gluon SD equation (2.1)}


The simplest way to show up the previous solution to the gluon SD equation (2.1) is 
to put  $\Delta^2_t(D) = 0$ and  $\xi=\xi_0 =\lambda^{-1}$ (due to the relations (3.11), of course) in the system of eqs.~(5.5)-(5.6), then it looks like

\begin{equation} 
D^{PT}_{\mu\nu}(q) =  { i T_{\mu\nu}(q)  \over q^2[ 1 + \Pi^s(q^2; D^{PT})] } + i \lambda^{-1} L_{\mu\nu}(q) { 1 \over q^2},
\end{equation}
so that the gluon invariant function is $d_{PT}(q^2) =[ 1 + \Pi^s (q^2; D^{PT})]^{-1}$, where the function $\Pi^s (q^2; D^{PT})$ is regular at zero and having only logarithmic 
divergences at large $q^2$: otherwise remains arbitrary. Formally putting $\Pi^s(q^2; D^{PT}) =0$ by hand, one obtains the free gluon propagator (3.3). 
The ST identities in case are

\begin{equation}
q_{\mu}q_{\nu} D^{PT}_{\mu\nu}(q) =q_{\mu}q_{\nu} D^0_{\mu\nu}(q) = i \lambda^{-1}.
\end{equation}
The corresponding gluon SD eq.~(5.1) now becomes

\begin{equation}
\hspace{-10mm} D^{PT}_{\mu\nu}(q) = D^0_{\mu\nu}(q) + D^0_{\mu\rho}(q)i T_{\rho\sigma}(q) q^2 \Pi^s(q^2; D^{PT}) D^{PT}_{\sigma\nu}(q).
\end{equation}

From above and now on in this section we denote the corresponding full gluon propagator $D$ by $D^{PT}$ (for the explanation see the text below). 
Obviously, the expression for the PT full gluon propagator (7.1), being free from all the kind of a mass scale parameters, describes the propagation of the PT massless gluons, since it has the PT singularity on the mass-shell $q^2=0$, i.e., the singularity of the free gluon propagator $\sim 1/q^2$.  It does not provide any hint how to prevent the free and  the PT massless gluon states to appear at large distance 
($q^2 \rightarrow 0$), i.e., it is not confining. The equality $\xi =\xi_0 =\lambda^{-1}$ takes place only for the regularized massless gluon fields. 

The system of the gluon SD equations of motion (7.1)-(7.3) is a well-elaborated system, see the most recent review~\cite{35} (and references therein). 
In this system the gauge symmetries of the QCD Lagrangian and its ground state coincides, like in QED~\cite{7,13,27} (if one ignores all the YM fields in our derivations, then the QCD 
will becomes nothing else but like QED). That is why we can conventionally call it as the PT system of the gluon SD equations (7.1)-(7.3), though the coupling constant remains strong, apart from the AF regime.
This system is not our concern. We present it here within our formalism for the readers convenience to directly compare it with the new solutions to the gluon SD equation (5.1)  
and its expressions presented in the relations (5.5)-(5.6). In order to reproduce in this case the system of the relations analogous to the system of the relations 
(5.7)-(5.8), it is necessary to put there $\Delta^2_t(D) = 0$, then  $\xi =\xi_0 =\lambda^{-1}$, as it has to be. Let us note that the first equalities in the relations (5.8) are valid for both solutions. 

The dimensional regularization method (DRM)~\cite{4,6,36,37,38} provides a gauge-invariant scheme to correctly calculate 
the finite parts of the QD loop integrals, while omitting their QD constants, i.e., simply to ignore them and to deal further only with the logarithmic divergences of the PT. 
This was a prescription rather than an exact result, as pointed out in~\cite{4}. From now on this prescription has been put on a firm mathematical ground 
(see the relations (4.9) and (5.8) and remarks just above).


\section{Conclusions}


The new solutions to the gluon SD equation of motion derived in section 5 requires the explicit presence of the mass squared scale parameter - the tadpole term - in the QCD 
ground state, i.e., at the fundamental quark-gluon level, as this was exactly proven  in the corresponding sections. 
In its presence the role of the QCD coupling constant $g^2$ becomes unimportant. This is also evidence of the 'dimensional transmutation',
$g^2 \rightarrow \Delta^2_t(D)$~\cite{2,39,40}, which occurs whenever a massless theory acquires mass dynamically. It is a general feature of spontaneously
symmetry breaking in field theories. We distinguish between the previous and new solutions to the QCD gluon SD equation (2.1) by the explicit presence of the tadpole term in the latter one, 
and not by the magnitude of the coupling constant even at the regularized gluon fields level yet. In both cases the gluon fields remain strongly interacted, apart from the AF regime~\cite{11,12}.
 
The gauge symmetry of the QCD Lagrangian is spontaneously broken in its ground state by the explicit presence of the tadpole term there. It is dynamically generated by the self-interaction 
of the multiplying massless gluon modes, involving the point-like four-gluon vertex only~\cite{10, 13, 41}, i.e., it is a fundamental quantity (''mass without mass''~\cite{42}), so that it cannot be a bound 
state of anything. This contribution to the full gluon self-energy does not depend on the external gluon momentum, and thus generates the scale parameter, having the dimension of a mass squared~\cite{6}, indeed. Its renormalization version has been conventionally called a mass gap in accordance with the question asked in the Introduction and sheds light on understanding ''one of the deep unsolved physics 
mysteries, the mass gap''~\cite{10} (see discussion below). The QCD is a self-consistent quantum gauge field theory. It does not need any 
extra degrees of freedom (such as Higgs fields~\cite{43}) or to extract a mass by some other ways at the quark-gluon level.
The mass gap cannot be disregarded from the theory and its ground state by any means, but the PT renormalizability of QCD will not be affected. 

The exact constraint (3.9) on any solution to QCD has been obtained before any proper subtraction scheme 
could have been formulated, and no any approximations/truncations/assumptions have been made in its derivation. 
So that all the other derivations, theorems, results and so on have to be agreed with it and {\it not \ vice \ versa}. 
It leads to the discovery of a new solution to the gluon SD eq.~ (2.1) in section 5, coinciding with the previous solution in section 7, at high energies, and 
different from it at finite and low energies.

The most important theoretical result after the formulation of QCD~\cite{1} and discovery of its AF behaviour~\cite{11,12}, has been obtained 
by Jaffe and Witten (JW)~\cite{10}.  One of their Millennium Prize Problems reads as follows:

\hspace{0.1mm}

{\bf JW's theorem: Yang-Mills existence and Mass Gap.} Prove that for any compact simple gauge group $G$,  a non-trivial quantum Yang-Mills theory exists  on $R^4$ and has a
mass gap $\Delta > 0$. Existence includes establishing some strong axiomatic properties for the corresponding Euclidean Green's functions.  

\hspace{0.1mm}

At the beginning of this century it became clear that the existing at that moment the QCD theory is not able to explain colour confinement and thus is not satisfying to the 
boundary condition in the IR. In their description of the theorem it is explained why a correct QCD theory must have the following three properties:
1). The nuclear force is strong but short-ranged due to the existence of the  ''mass gap'', a scale parameter $\Delta > 0$.
2). The physical particle states are $SU(3)$ - invariant,i.e., confinement of colour degrees of freedom at the fundamental quark-gluon level. 
3). The chiral symmetry breaking to account for the ''current algebra'' theory of soft pions.  

All our advance results obtained in this paper can be also formulated similar to JW's theorem as the following statement, namely

\hspace{0.1mm}

{\bf Mass Gap existence and Yang-Mills theory.} If a non-trivial quantum Yang-Mills theory with gauge group $SU(3)$ exists on $R^4$ then it has been proven that it should have a mass gap 
$\Delta^2 > 0$. The proof includes establishing the exact constraint (3.10) on any solution to QCD, depending on the mass gap, which dynamical origin is fixed. 
Such defined mass gap determines a scale of the confinement phase transition in QCD.

\hspace{0.1mm}   
   
The differences between the definitions of the JW's mass gap and ours is as follows: they define the mass gap as a solution of the Hamiltonian problem that  every excitation 
of the vacuum has energy at least  $\Delta > 0$. Such defined mass gap being in the physical spectrum can be responsible for the second PCAC phase transition at the hadronic level. 
But being defined in the physical spectrum it cannot explain the linear rising potential between heavy quarks at the quark-gluon level. For this we need the presence of a mass 
squared scale parameter in the confining solution(s) of the full gluon propagator. In other words, the confinement phase transition needs its own mass gap. 
To our mass gap  can be assigned a physical meaning as a scale parameter determining the strengths of  all the different NP effects and quantities in the true QCD vacuum 
(such as the string tension, scale breaking, etc., so that they have to be measured in terms of the mass gap $\Delta^2$). In this way the conception of the mass gap 
can and must be also extended to the first confinement phase transition. Otherwise it will be not possible to explain and calculate the linear rising potential between heavy quarks (mentioned above) 
as well as many other NP effects and quantities, having the dimensions of a mass gap in the different powers such as the vacuum energy density, the Bag constant, the condensates, 
the topological susceptibility, etc.~\cite{13} (and references therein).

Let us also take into account that the JW's theorem is formulated in the most general way, not specifying the meaning of the mas gap itself, which making it possible for its different interpretations.
As it follows from above, by the term ''mass gap'' we mean not only a mass of the physical or virtual particles, but also the different types of scale breaking etc.
No doubts that this theorem is perfectly fitted into the our results, formulated similar to the JW's theorem, and  {\it \ vice \ versa} they are perfectly fitted into the JW's theorem.
This interrelation increases the mathematical and physical contents power of both JW's theorem and our results.
In order that a correct QCD will may produce ''a mathematically complete example of quantum gauge field theory in four dimensional space-time''~\cite{10}, it is necessary, 
in addition, to calculate the confining full quark propagator within this theory, but this is beyond  the scope of the present series.
It has to be done in the separate paper as a subject for further work. Completing this work, one can think next how to prove the existence of a non-trivial QCD/YM theory, 
since the analytic and asymptotic properties of the corresponding Euclidean Green's functions will be known at that moment. It will make it possible to establish some axioms for them,
needed to prove JW's theorem itself?
   
Concluding finally, let us make a few general things perfectly clear. The tadpole term, even breaking the QCD Lagrangian gauge symmetry in its ground state, 
cannot be thrown out away from the gluon SD equation. Omitting any term/parameter in any NL transcendental equation, especially such complicate as the gluon equation of motion, one can 
lose an important  piece of information about  its possible solutions (the number of the solutions, their structures and properties, etc.). The tadpole term can be removed from 
the consideration only on the basis of the mathematical relation as it has been precisely described in this paper. The connection of our results to the well-known Elitzur's theorem~\cite{44} 
left out of the discussion here. Without knowing the confining solution(s) for the full gluon propagator (which will be present in the second and third papers of our series) such discussion will not 
be relevant. So that it will be present in the final paper.

\section*{Acknowledgments}

The authors are grateful to P. Forg\'{a}cs, J. Nyiri, T.S. Bir\'{o}, M. Vas\'{u}th, Gy. Kluge
for useful suggestions, remarks, discussions and help. The work was supported by the Hungarian National Research, Development and Innovation Office (NKFIH) under the contract 
numbers OTKA K135515 and NKFIH 2019-2.1.11-T\'ET-2019-00078, 2019-2.1.11-T\'ET-2019-00050, the Wigner Scientific Computing Laboratory.

\appendix


\section{The full gluon propagator up to one loop}


\begin{figure}[h!]
\begin{center}
\includegraphics[width=13.0truecm]{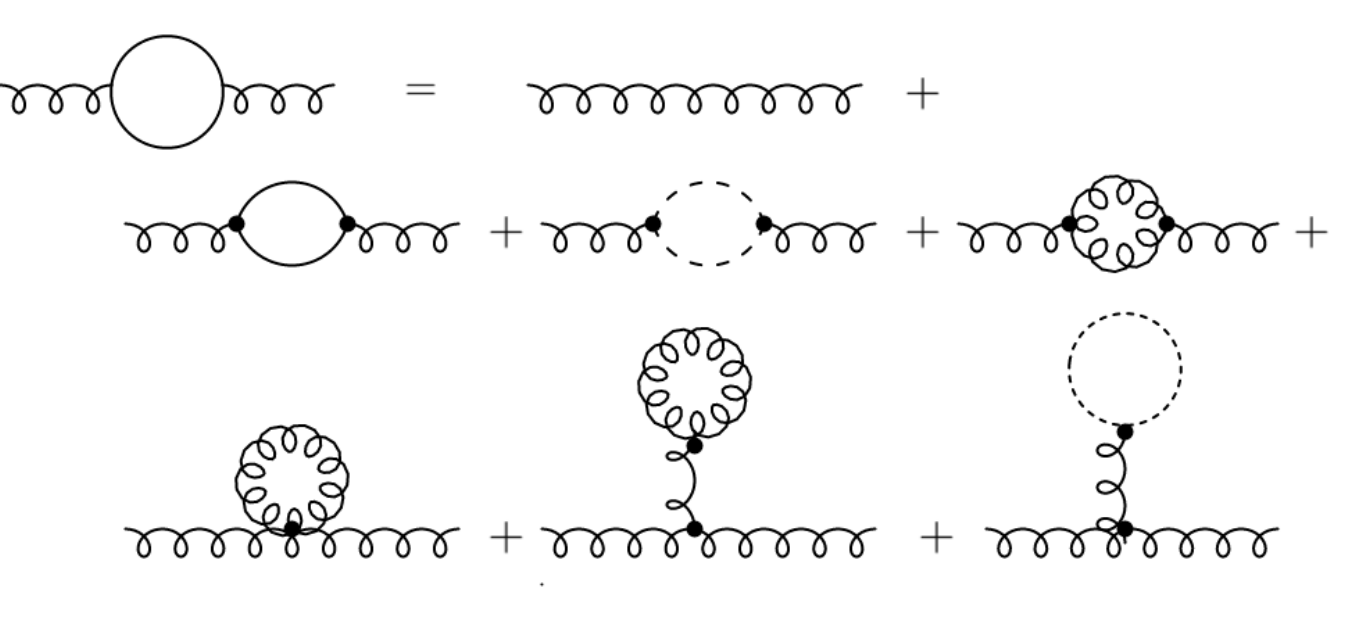}
\caption{The SD equation for the full gluon propagator up to one loop as present in~\cite{6}. It is useful to compare it with Fig. 1 and its description in the text after it.}
\label{fig:1}
\end{center}
\end{figure}

In order to explicitly compare the tadpole term (2.4) with the other tadpole-type terms 
it is instructive to consider the iteration of the gluon SD eq.~(2.1) up to one loop. It is shown in Fig. A.2. The last two terms in this figure appear because
of the NL nature of the iteration series when the next iterations contribute to the previous ones. That is the reason why such multiplied types
of the tadpole terms are not present in the skeleton loop decomposition for the full gluon propagator in Fig. 1, but have to be shown up in the iteration terms, seen in Fig. A.2. This iteration has been exactly calculated in~\cite{6} by using the DRM,
while rightfully now ignoring the tadpole term itself and all the other tadpole-likes ones, in accordance with the remarks at the end of section 7.
It is worth emphasizing that in~\cite{6} this iteration has not been called the PT expansion up to $O(g^2)$, though looks like it, but expansion up to $O(\hbar)$. 
It is well-known that in QCD the expansion in powers of the coupling constant does not make any sense because it
is strong. The PT expansion in QCD makes sense only in the AF~\cite{11,12} regime when it becomes weak.
In order to perform analytical derivations below, we will take the interaction vertices as the point-like ones. 
The external gluon propagators will be taken as the free ones. All the skeleton loop propagators will remain as the full ones. Then the corresponding iteration will look like Fig. A.2, and 
hence can be called as quasi-one loop expansion.

The tadpole term up to one skeleton loop, shown first in the third line of Fig. A.2, analytically can be written down as follows:

\begin{equation}
\Pi_{\rho\sigma}^t(D) \sim \int d^4 l T^0_{\rho\sigma\alpha\beta}D_{\alpha\beta}(l) = \delta_{\rho\sigma} \Delta^2_t(D),
\end{equation}
where and everywhere below all the overall numerical numbers and the colour group factors will be omitted, since they are not important for our purpose.
Being the QD but already regularized constant, it is not connected to any external gluon momentum.
In this respect it is different from all other tadpole-like terms which appear as a result of the subtraction scheme or in the NL iteration procedure.
Also it directly generate the mass squared parameter, and thus explicitly violets the gauge symmetry of the QCD Lagrangian in its ground state~\cite{6}.
We have shown that in the PT QCD it should be neglected on the general basis as well as all other tadpole-type terms. But in the NP QCD it should remain intact, while all other 
tadpole-types terms have to be removed from the theory, as it has been proven in section 4.

Let us now investigate the tadpole-like term with the gluon skeleton loop, shown second in the third line of Fig. A.2. Its analytical expression is 

\begin{equation}
\Pi^{(4)}_{\rho\sigma}(q,p) \sim T^0_{\rho\sigma\mu'}(q,p) D^0_{\mu\mu'} (p) \int d^4 l \ T^0_{\nu\zeta\mu}(l,p) D_{\nu\zeta} (l),
\end{equation}
where the gluon momentum $p$ is, in fact, zero, i.e., $p=0$ but it is convenient to go to this limit at the final step only.
Both gluon propagators in the t' Hooft-Feynman gauge are $D^0_{\mu\mu'} (p)= \delta_{\mu\mu'} / p^2$ and
$D_{\nu\zeta} (l)= \delta_{\nu\zeta} d_g(l^2)/ l^2$, where $d_g(l^2)$ is the corresponding invariant function of the full gluon propagator.
Let us remind that such kind of the skeleton loop integrals are assumed to be regularized from above and below.
Using now the Euclidean space Feynman rules for the corresponding vertices present in~\cite{6}, one arrives at

\begin{eqnarray}
T^0_{\rho\sigma\mu'}(q,p) &=& -( 2q+p)_{\mu'} \delta_{\rho\sigma} + (q - p)_{\sigma}\delta_{\rho\mu'} + (2p + q)_{\rho}\delta_{\sigma\mu'}, \nonumber\\
T^0_{\nu\zeta\mu}(l,p) &=& 2(l+p)_{\mu} \delta _{\nu\zeta} - l_{\zeta}\delta _{\nu\mu} - (2p + l)_{\nu}\delta _{\zeta\mu}.
\end{eqnarray}
Substituting thee expressions into the eq.~(A.2), one finally obtains

\begin{equation}
\Pi^{(4)}_{\rho\sigma}(q,p)  \sim T^0_{\rho\sigma\mu}(q,p)  {2  \over p^2} \int d^4 l { d_g(l^2) \over l^2} [ 3l_{\mu} + p_{\mu}].
\end{equation}

By the symmetry integration ($d^4l =l^3 dl = (1/2)l^2dl^2$ and all the overall numerical numbers due to the integration over angular variables in four-dimensional Euclidean space~\cite{6} omitted) 
the first loop integral is zero, while the second one leads to

\begin{equation}
\Pi^{(4)}_{\rho\sigma}(q,p)  \sim T^0_{\rho\sigma\mu}(q,p) p_{\mu} {2 \over p^2} \times \tilde{\Delta}^2_g(d),
\end{equation}
where

\begin{equation}
\tilde{\Delta}^2_g(d) = \int d^4 l { d_g(l^2) \over l^2} = 0.
\end{equation}
This constant is the QD but regularized skeleton loop integral, having the dimension of mass squared. It is one of the constants present in the
relations (5.7)-(5.8).  It should be discarded on the general basis within our approach, i.e., put formally zero,
even before the final $p^2 = - M^2 \rightarrow 0$ limit, as already shown above. Let us note that the cancellation of the pole
$1/p^2 = - 1/ M^2$ in eq.~(A.5) can be always achieved by going to the dimensionless variable $x= l/M$ in the skeleton loop integral
(A.6), i.e., this pole is not a problem here. So that $\Pi^{(4)}_{\rho\sigma}(q,p) =0$ because of eq.~(A.6).

Let us finally investigate the tadpole-like term with the ghost skeleton loop, shown third in the third line of Fig. A.2. In analogy with (A.2) its analytical expression is 

\begin{equation}
\Pi^{(5)}_{\rho\sigma}(q,p) \sim T^0_{\rho\sigma\mu'}(q,p) D^0_{\mu\mu'} (p) \int d^4 l { l_{\mu} \over l^2} d_{gh}(l^2),
\end{equation}
where $d_{gh}(l^2)$ is the invariant function for the full ghost propagator $G(l)$.
The vertex $T^0_{\rho\sigma\mu'}(q,p)$ is explicitly shown in the relations (A.3), and the remaining free gluon propagator is also the same.
The altered rule~\cite{6} for the gluon-ghost vertex gives $(1/2)[(l_{\mu}+ l{\mu})=l_{\mu}$ and the gluon momentum $p$ is zero again at the final stage.
Since the nominator of this skeleton loop integral contains the loop variable linearly, then by the symmetry integration it is simply zero from the very beginning, i.e.,
$\Pi^{(5)}_{\rho\sigma}(q,p) = 0$.

Concluding, neither the tadpole-like term with the skeleton gluon loop nor its ghost counterpart contribute into the full gluon propagator within our
approach. In these cases the subtractions with respect to the external gluon momentum $q$ (though possible), but were not even necessary
to perform, since the final results would have been the same zeros.
The more severe IR structure (than the free gluon propagator has) of the full gluon propagator are to be treated within the distribution
theory~\cite{30} as underlined above, and is beyond the scope of the present work.  The one of the primary goal of this work is to prove
that the mass gap exists in the QCD/YM theory. So its dynamical source in the ground state - the tadpole term - cannot be ignored 
by any means. All the other tadpole-like terms are to be always ignored on the general basis, i.e., put zeros,  as explicitly demonstrated here as well.

\hspace{5mm}

\end{document}